\documentclass[%
 reprint,
 superscriptaddress,
footinbib,
nobibnotes,
 amsmath,amssymb,
 aps,
 prl,
]{revtex4-2}

\usepackage{graphicx}
\usepackage{dcolumn}
\usepackage{hyperref}
\usepackage{newtx}

\begin{document}

\title{Harmonic and Subharmonic Magnon Generation in a Surface Acoustic Wave Resonator}

\author{Yunyoung Hwang}
\altaffiliation[]{These authors contributed equally to this work.}
\affiliation{%
 CEMS, RIKEN, 2-1, Hirosawa, Wako 351-0198, Japan
}%
\affiliation{%
 Institute for Solid State Physics, University of Tokyo, Kashiwa 277-8581, Japan
}%

\author{Liyang Liao}
\altaffiliation[]{These authors contributed equally to this work.}
\affiliation{%
 Institute for Solid State Physics, University of Tokyo, Kashiwa 277-8581, Japan
}%

\author{Jorge Puebla}%
 \email{puebla.jorge.8m@kyoto-u.ac.jp}
\affiliation{%
 CEMS, RIKEN, 2-1, Hirosawa, Wako 351-0198, Japan
}%
\affiliation{%
 Department of Electronic Science and Engineering, Kyoto University, Kyoto 615-8510, Japan
}%

\author{Marco Br\"uhlmann}
\affiliation{%
 Institute of Theoretical Physics and Vienna Center for Quantum Science and Technology, TU Wien, A-1040 Wien, Austria
}%

\author{Carlos Gonzalez-Ballestero}
\affiliation{%
 Institute of Theoretical Physics and Vienna Center for Quantum Science and Technology, TU Wien, A-1040 Wien, Austria
}%

\author{Kouta Kondou}
\affiliation{%
 CEMS, RIKEN, 2-1, Hirosawa, Wako 351-0198, Japan
}%

\author{Naoki Ogawa}
\affiliation{%
 CEMS, RIKEN, 2-1, Hirosawa, Wako 351-0198, Japan
}%
\affiliation{%
 Tokyo College, University of Tokyo, Tokyo 113-8656, Japan
}%

\author{Sadamichi Maekawa}
\affiliation{%
 CEMS, RIKEN, 2-1, Hirosawa, Wako 351-0198, Japan
}%
\affiliation{%
 Advanced Science Research Center, Japan Atomic Energy Agency, Tokai 319-1195, Japan
}%

\author{Yoshichika Otani}
\email{yotani@issp.u-tokyo.ac.jp}
\affiliation{%
 Institute for Solid State Physics, University of Tokyo, Kashiwa 277-8581, Japan
}%
\affiliation{%
 CEMS, RIKEN, 2-1, Hirosawa, Wako 351-0198, Japan
}%

\date{\today}

\begin{abstract}
We experimentally observe the generation of magnon harmonics and subharmonics in an on-chip surface acoustic wave resonator incorporating a thin Co$_{20}$Fe$_{60}$B$_{20}$ film, using micro-focused Brillouin light scattering.
In our devices, rotating the in-plane magnetic field allows continuous tuning of the magnon–phonon coupling from weak to strong within the same resonator.
In the weak coupling regime, we only observe fundamental magnetoelastic wave signal at $f_{1}$. 
Conversely, in the strong coupling regime, in addition to the fundamental magnetoelastic wave, we observe subharmonic and harmonic signals at $3/2f_{1}$, $2f_{1}$, and $3f_{1}$, which are well reproduced by our analytical model.
Our results establish phonons as a means to generate and control nonlinear magnons in the strong coupling regime, providing a new route for magnonic signal processing.
\end{abstract}


\maketitle
Hybrid systems of magnons (quanta of spin
waves) and phonons (quanta of acoustic waves) provide a solid-state platform for coherent transduction and control of spin dynamics~\cite{magnon-phonon_review,Jorge_perspective_APL}.
While the linear magnon–phonon coupling is well explored, accessing and controlling nonlinear processes---such as parametric generation of magnons---are desired~\cite{HwangJPS,Geilen2022,Liu2024,Matsumoto2024,Geilen2025}.
This is because nonlinear phenomena in coupled oscillator systems are a rich subject across physics, from Josephson parametric amplifiers in superconducting circuits~\cite{Josephson1,Josephson2} to optical parametric down-converters in photonics~\cite{Lounis2005}.
However, those nonlinear operations typically require stringent microwave-pump conditions~\cite{Josephson_Parametric}.
A central question is whether magnetoelastic coupling can be used to drive and isolate such nonlinear spin-wave phenomena in a simple, tunable on-chip setting.

In this Letter, we demonstrate a switch process of nonlinear magnon generation in a thin Co$_{20}$Fe$_{60}$B$_{20}$ film constructed in a two-port surface acoustic wave (SAW) resonator by rotating the in-plane magnetic field at room temperature.
In our device, when the in-plane magnetic field angle relative to the SAW propagation ($\phi_H$) is $0^{\circ}$, strong magnon--phonon coupling---where the coupling strength exceeds both the magnon and phonon damping~\cite{Kockum2019}---is observed at the resonant frequency $f_1$.
Conversely, weak coupling is observed when $\phi_H = 30^{\circ}$~\cite{HwangPRL}.
This is a consequence of the strong angle dependence of magnetoelastic coupling by shear-horizontal SAWs, which is crucial for realizing robust magnon–phonon coupling~\cite{HwangPRL,Babu2021,SAWrabi,NTTarxiv}.
Under strong coupling we observe (i) accompanying harmonics at $2f_1$ and $3f_1$ at modest drive powers; and
(ii) a fractional spectral line at $f_{3/2}$ that signifies one-phonon $\rightarrow$ two-magnon parametric down-conversion (magnon splitting) by the micro-focused Brillouin light scattering (BLS)~\cite{BLS}.
Note that tuning the coupling strength from weak to strong coupling directly affects the efficiency of transferring the driving input to the coupled waves; in our case, transferring the SAW driving to the excitation of magnetoelastic waves.
An analytical model reproduces the main features and their field-angle dependence~\cite{Theory}.
This platform allows us to acoustically trigger nonlinear generation of magnons on demand by just changing the orientation of the field.

\begin{figure}[]
\centering
\includegraphics{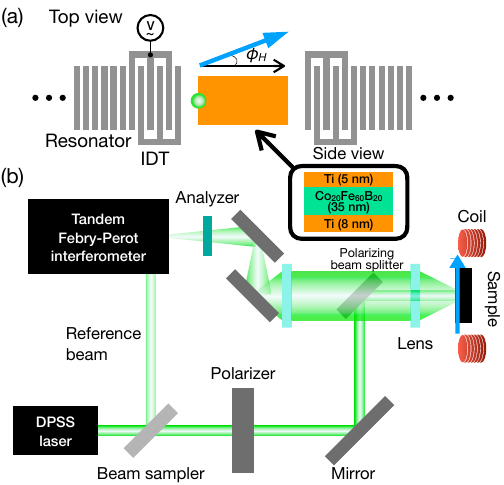}
\caption{\label{fig1} (Color online)
(a) Schematic illustration of the SAW resonator device including a Ti/Co$_{20}$Fe$_{60}$B$_{20}$/Ti heterostructure.
A signal generator applies RF power to one of the interdigital transducers (IDTs). The BLS reference laser illuminates the side close to the exciting IDT.
During the measurements, an external in-plane magnetic field is applied at an angle $\phi_H$ relative to the SAW propagation direction.
(b) Schematic of the micro-focused BLS measurement setup.}
\end{figure}

To observe the effect of magnon--phonon coupling strength on nonlinearities, we use two-port SAW resonator devices incorporating a Ti(8 nm)/Co$_{20}$Fe$_{60}$B$_{20}$(35 nm)/Ti(5 nm) structure on a 128$^{\circ}$ Y-cut LiNbO$_{3}$ substrate, as shown in Fig.~\ref{fig1}(a).
Details on the device fabrication method and procedure can be found in Ref.~\cite{HwangPRL}.
It is important to note that the SAW generated by an interdigital transducer (IDT) propagates along the crystal’s $x$-axis in LiNbO$_3$.
In our previous reports, we characterized our devices by monitoring the in-plane magnetic field dependence of SAW phonon transmissions.
Here we utilize a well-known spectroscopy method to detect magnons, the micro-focused Brillouin light scattering (BLS)~\cite{BLS,Geilen2022,Geilen2025}.
A schematic of the BLS measurement setup is displayed in Fig.\ref{fig1}(b).
For more details on the BLS technique, see Refs.~\cite{BMLS, BLS}.
 
\begin{figure}[]
\centering
\includegraphics[scale=1]{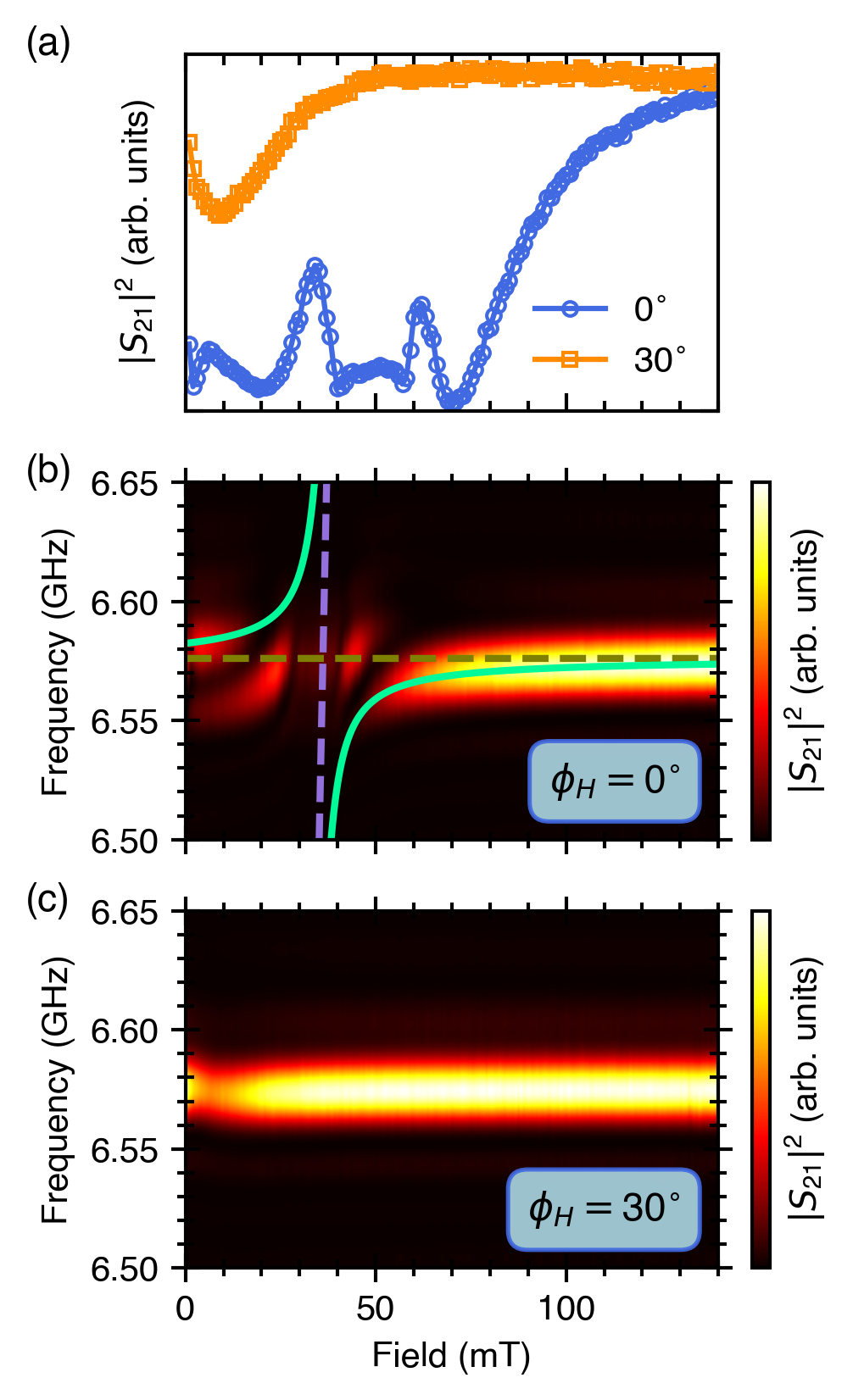}
\caption{\label{fig2} (Color online)
(a) Measured SAW transmission signal $|S_{21}|^2$ by VNA at the frequency of 6.57 GHz as a function of the amplitude of the externally applied in-plane magnetic field
The in-plane field angles ($\phi_H$) of each curve are shown in the legend.
At $\phi_H=0^{\circ}$ (orange open squares), there are two distinct absorption dips, while only one absorption dip is recognizable at $\phi_H=30^{\circ}$ (blue open circles).
(b,c) Measured $|S_{21}|^2$ by VNA as a function of the frequency and the amplitude of the external in-plane magnetic field when (b) $\phi_H = 0^{\circ}$ and (c) $\phi_H =30^{\circ}$.
The green solid curves, purple dashed line, and khaki dashed line show the fitting of magnon--phonon anticrossing, magnon frequency, and phonon frequency, respectively.}
\end{figure}

As shown in Fig. \ref{fig1}(a), we used a SAW device with a 35-nm-thick Co$_{20}$Fe$_{60}$B$_{20}$ layer.
One key feature of our hybrid magnon--phonon SAW devices is the ability to transition between weak and strong coupling regimes simply by adjusting the in-plane magnetic field angle ($\phi_H$) relative to the SAW propagation direction.
We show this key feature in Fig. \ref{fig2}(a) through the SAW transmission signal ($|S_{21}|^2$) measured by a vector network analyzer (VNA) at the resonant frequency, prior to presenting the BLS results.
The IDT at the left and right side in Fig.~\ref{fig1}(a) are connected to port1 and port2 of the VNA, respectively.
We apply an in-plane magnetic field with a magnitude of 200 mT to saturate the magnetization and measure $|S_{21}|^2$ while varying the field magnitude with a step of 1 mT.
In the strong coupling regime, we observe two absorption dips as a function of the in-plane external field, as shown in Fig.~\ref{fig2}(a) for $\phi_H=0^{\circ}$.
Conversely, the weak coupling regime manifests as a single absorption dip as a function of the in-plane external field, as shown in Fig.~\ref{fig2}(a) for $\phi_H=30^{\circ}$.
This distinction arises from phonon frequency shifts resulting from the magnon--phonon band splitting or anticrossing in our device, a typical consequence of strong coupling~\cite{Kockum2019,SAWrabi,NTTarxiv}.
When $\phi_H = 0^{\circ}$, SAW resonant frequency shows clear frequency shifts forming anticrossing [Fig.~\ref{fig2}(b)],
while no anticrossing has been observed when $\phi_H=30^{\circ}$ [Fig.~\ref{fig2}(c)].
Fitting the anticrossing as described in Ref.~\cite{HwangPRL} [green curves in Fig.~\ref{fig2}(b)] yields a magnon--phonon coupling strength of $g=2\pi\times150$ MHz.
The magnon and phonon relaxation rates are $\kappa_m = 2\pi\times53$ MHz and $\kappa_p= 2\pi\times16$ MHz, respectively~\cite{HwangPRL}.
These parameters ensure the conditions for the strong coupling regime $(g > \kappa_m, \kappa_p)$, as demonstrated in our previous work~\cite{HwangPRL}.

\begin{figure}[]
\centering
\includegraphics[scale=1]{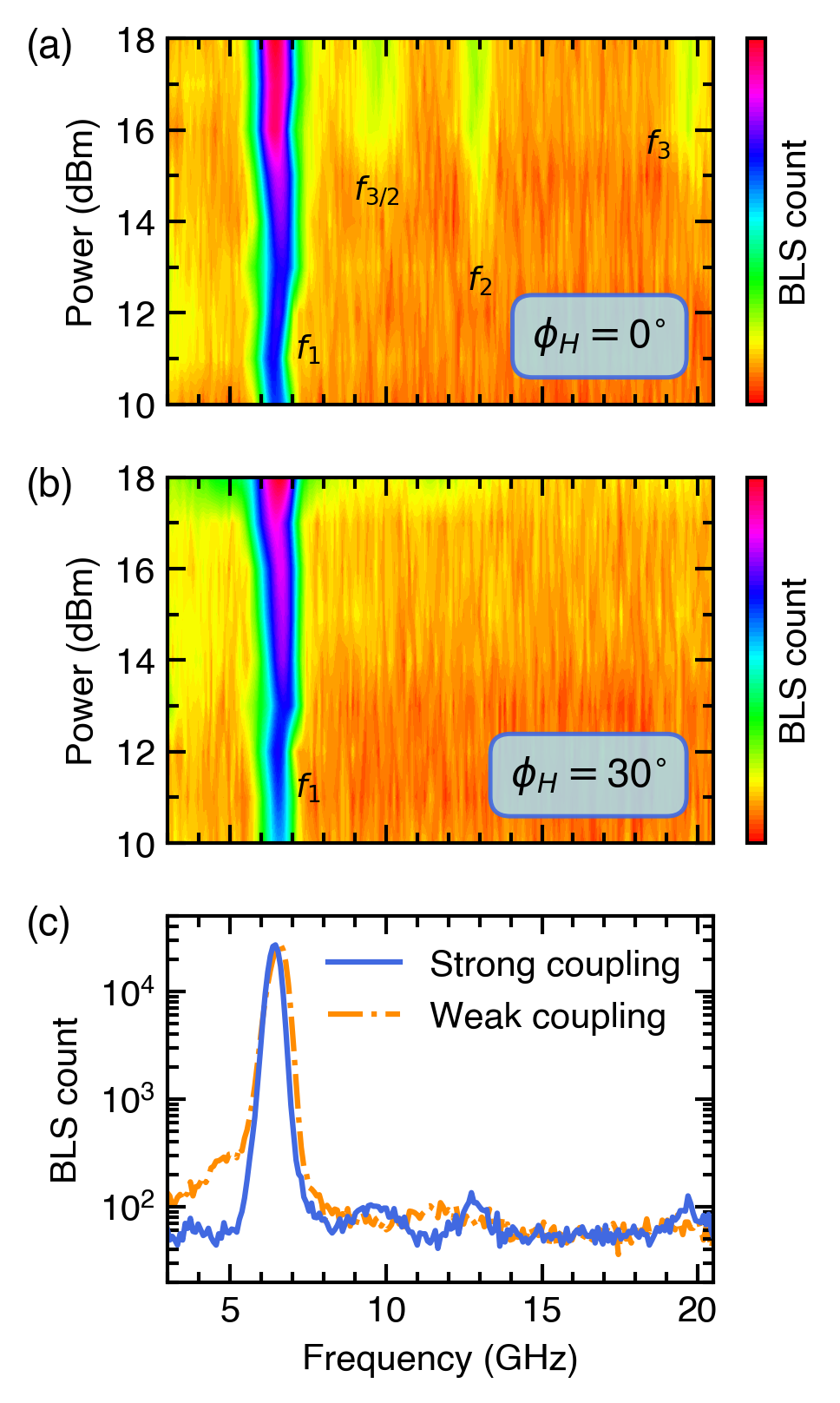}
\caption{\label{fig3} (Color online)
(a,b) Dependence of BLS signal on the driving power of the fundamental SAW excitation at $f_{\mathrm{SAW}}=f_{\mathrm{1}}$, under the in-plane magnetic field with (a) $\mu_0 H = 36$ mT and $\phi_H=0^{\circ}$ (strong coupling), and (b) $\mu_0 H = 9$ mT and $\phi_H=30^{\circ}$ (weak coupling).
(c) Individual BLS spectra recorded at a driving power of 16 dBm under strong coupling (solid blue line) and 18 dBm under weak coupling (dashed orange line).}
\end{figure}

Now we come to the central result of our study.
Figure~\ref{fig3}(a) shows the dependence of the BLS signal on the driving power of the fundamental SAW excitation at $f_{\mathrm{SAW}}=f_{\mathrm{1}}$, with the in-plane magnetic field fixed to the resonance condition attained from Fig.~\ref{fig2}(b); the magnitude of $\mu_0 H = 36$ mT and $\phi_H=0^{\circ}$ (strong coupling).
The driving power was systematically increased from 10 dBm to 18 dBm.
Around 15 dBm, in addition to the fundamental BLS signal at $f_{\mathrm{1}}$, a signal  $f_{\mathrm{2}} = 2f_1$ emerges.
At higher powers, around 16 dBm, we observe the appearance of signals $f_{\mathrm{3}} = 3f_1$ and $f_{\mathrm{3/2}} = (3/2)f_1$.
In contrast, Fig.~\ref{fig3}(b) illustrates the BLS signal's power dependence when $\mu_0 H = 9$ mT and $\phi_H = 30^{\circ}$ (weak coupling).
Unlike Fig.~\ref{fig3}(a), no BLS signals corresponding to $f_{\mathrm{3/2}}$, $f_{\mathrm{2}}$, or $f_{\mathrm{3}}$ in Fig.~\ref{fig3}(b) are observed in this case.
To further illustrate the comparison, Fig.~\ref{fig3}(c) presents individual BLS spectra at a driving power of 16 dBm under strong coupling (solid blue line) and 18 dBm under weak coupling (dashed orange line).
Despite similar BLS counts for the fundamental excitation at $f_{\mathrm{1}}$, the magnetoleastic instabilities, corresponding to excitations at $f_{\mathrm{3/2}}$, $f_{\mathrm{2}}$ and $f_{\mathrm{3}}$, are only observable under the strong coupling condition.   

\begin{figure}[]
\centering
\includegraphics[]{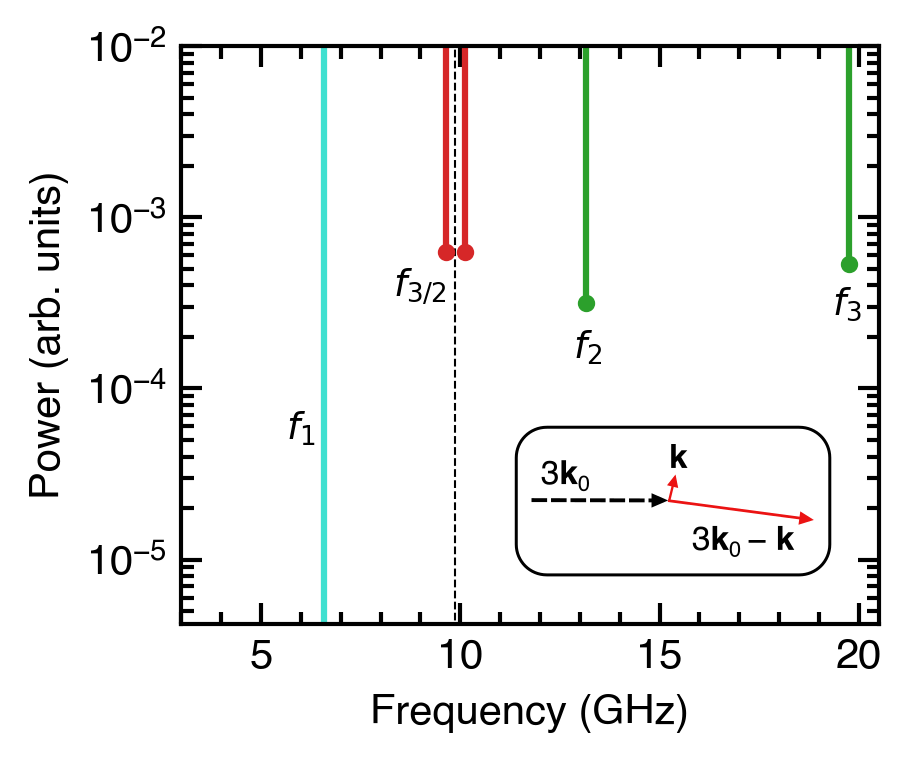}
\caption{\label{fig4} (Color online)
Predicted frequencies and threshold driving powers for the first parametric process (phonon splitting into two magnons, red) and for higher harmonics (green) to rise above the noise floor at $f_2$ and $f_3$.
The fundamental driving frequency $f_1$, and its 3/2 multiplied frequency are illustrated by the cyan solid line and the black dashed line, respectively.
The inset shows the relative wavevectors and propagation angle giving rise to incoming phonon and outgoing magnons.}
\end{figure}

Through the establishment of a model for nonlinear magnon excitation in the presence of linear and nonlinear magnon-phonon interaction, we successfully replicate the excitation of both harmonic and subharmonic signals in Fig.~\ref{fig3}(a)~\cite{Theory}.
Here, we describe the mechanisms of nonlinear magnon generation predicted by our model.
By the driving SAW field at frequency $f_1$ and wavevector $\mathbf{k}_0$, phonons create magnons at the same frequency and wavevector due to the (linear) strong coupling.
Due to both the nonlinear magnon--phonon interaction and the nonlinear (3- and 4-) magnon--magnon interaction, higher magnon harmonics at $2\mathbf{k}_0$ and $3\mathbf{k}_0$ are excited.
Phonon harmonics at $2\mathbf{k}_0$ and $3\mathbf{k}_0$ are also excited by these processes as well as by the linear magnon-phonon coupling.
The predicted threshold powers for the harmonic magnon signals to appear above the noise are shown as green points in Fig.~\ref{fig4}.
Then, the $3\mathbf{k}_0$ phonons are annihilated and parametrically amplified magnon modes with the wavevectors $\mathbf{k}$ and $3\mathbf{k}_0-\mathbf{k}$ are created, as shown in the inset of Fig.~\ref{fig4}.
The $\mathbf{k}$ values and the corresponding frequencies and threshold powers of the two created magnons are numerically calculated and shown as red points in Fig.~\ref{fig4}.
It is important to note that once the $3\mathbf{k}_0$ magnon and the $3\mathbf{k}_0$ phonon are populated, there are many processes that can happen to generate parametric instability:
a $3\mathbf{k}_0$ magnon could split into two (3-magnon process); two $3\mathbf{k}_0$ magnons could split into two (4-magnon process);
or a $3\mathbf{k}_0$ phonon can split into two magnons.
We have calculated the threshold for each process and concluded that the latter process appears at a lower threshold power compared to the 3- and 4-magnon processes~\cite{Theory}.
This is a feature only enabled by our structure.
In other structures, magnon--phonon couplings would be smaller and one would observe 3-magnon and 4-magnon processes first.
The detailed description of the model and calculations can be found in 
Ref.~\cite{Theory}.
Note that, according to our model, the nonlinear signals at $f_{3/2}$, $f_2$, and $f_3$ do not appear when $\phi_H = 30^{\circ}$ due to not enough population of the $\mathbf{k}_0$ magnons, as experimentally confirmed [Fig.~\ref{fig3}(b)].

In summary, we have experimentally demonstrated that strong magnon--phonon coupling in a SAW-driven device can be leveraged to generate nonlinear magnonic signals on-chip at room temperature. 
By tuning the magnetic field angle, we turn on or off the emergence of subharmonic (3/2) and harmonic (2nd, 3rd order) spin wave excitations not observed in previous SAW--magnon studies lacking strong coupling~\cite{Weiler2011,Weiler2012,Dreher2012,MingranPRB,MingranSciAdv,HwangAPL,HatanakaPRAppl,HwangAMI,Geilen2022,HwangJPS,TomPRL,Geilen2025}.
Our results highlight how enhanced nonlinear dynamics arise from the efficient energy transfer in the strong coupling regime, providing a pathway to realize magnetoacoustic parametric amplification without any external microwave field.
This work expands the toolkit for magnon-based information processing~\cite{magnoncomputing}, suggesting new strategies for on-chip spintronic devices~\cite{HIROHATA2020166711,CHEN2023193} and magnonic neuromorphic computing architectures~\cite{Hughes2019,Papp2021,Dmytro2023}.
We anticipate that these findings will motivate further exploration of nonlinear magnon--phonon hybrids, for example by exploiting additional degrees of freedom (such as material anisotropies or temperature control) to engineer tunable nonlinear responses for advanced functional devices.

\begin{acknowledgments}
This work was supported by Grants-in-Aid for Scientific Research (S) (No. 19H05629) and the Japan Society for the Promotion of Science Grants-in-Aid for Scientific Research (No. 20H01865). Y.H. thanks to RIKEN Junior Research Associate Program for supporting this work. L.L. would like to thank the support from JSPS through Research Program for Young Scientists (No. 23KJ0778). J.P. and S.M. acknowledges support of JSPS KAKENHI No. 24K00576 from MEXT, Japan.
C.G.B. and M.B. acknowledge the Austrian Science Fund FWF for the support with the project PAT-1177623 ``Nanophotonics-inspired quantum magnonics.''
Y.O. acknowledges The LANEF Chair of Excellence and QSPIN project at University Grenoble Alpes.
Y.O. acknowledges support from JSPS KAKENHI Grant-in-Aid for Transformative Research Areas (A) “Chimera Quasiparticles for Novel Condensed-Matter Science” (No. 24H02233).
\end{acknowledgments}

\bibliography{manuscript.bbl}

\end{document}